\documentclass[submission,copyright,creativecommons]{eptcs}

\usepackage{iftex}

\ifpdf
  \usepackage{underscore}         
  \usepackage[T1]{fontenc}        
\else
  \usepackage{breakurl}           
\fi

\usepackage{xurl}
\usepackage{amsmath}
\usepackage{graphicx}
\usepackage{multirow}
\usepackage{xspace}

\usepackage[dvipsnames]{xcolor}
\hypersetup{colorlinks, breaklinks, urlcolor=Maroon, linkcolor=Maroon} 

\begin{document}

\newcommand{\chessdb}[0]{\textit{Chess\_db}\xspace}
\newcommand{\Chessdb}[0]{\textit{Chess\_db}\xspace}
\newcommand{\swi}[0]{\textit{SWI-Prolog}\xspace}
\newcommand{\prolog}[0]{\textit{Prolog}\xspace}
\newcommand{\pgn}[0]{\textit{PGN}\xspace}
\newcommand{\lichess}[0]{\textit{Lichess}\xspace}
\newcommand{\elite}[0]{\textit{Elite}\xspace}
\newcommand{\rocksdb}[0]{\textit{RocksDB}\xspace}

\newcommand{\pred}[2]{\textit{#1/#2}\xspace}

\newcommand{\sw}[1]{\textit{#1}\xspace}
\newcommand{\fixme}[1]{\textbf{fixme}: #1}
\title{\Chessdb: A framework for working with large chess game datasets.}
\author{Nicos Angelopoulos
\institute{University College \& Imperial College}
\institute{London, UK}
\email{nicos.angelopoulos@imperial.ac.uk}
\and
Jan Wielemaker
\institute{SWI-Prolog solutions}
\institute{Amsterdam, NL}
\email{jan@swi-prolog.org}
}

\def\titlerunning{\chessdb}
\def\authorrunning{N. Angelopoulos \& J. Wielemaker}


\maketitle

\begin{abstract}
Chess is a two player strategic game that is embedded in classical AI 
culture as it was once the frontier for intelligent behaviour. 
There was the silent assumption that the advent of computer engines that 
play better than the best humans will extinguish interest in the game.
However, the opposite has come to pass, with a growing following for the game.
A lot of the computational resources are now centered around training of
players, where the engine output is just one aspect. Access to 
past games is also an essential part, both in knowing what games a specific player
has played previously, and also which continuations at a certain position
have led to victory more often for each of the two colour players.
We present \chessdb a suite of logic programming tools that can effectively manipulate
games both in memory and via creating back end databases. 
In particular, we provide versatile code that creates databases from PGN (portable game notation)
game files and explore the suitability of open source key-value databases for 
storing position tables that provide near-instant access to 
information pertaining to substantially large number of games.
\end{abstract}


\section{Introduction}

Chess has a long association with classical AI and was once the frontier of 
algorithmic development in the field. Despite the development of engines running 
on widely available, inexpensive computational platforms such as mobile phones that are better than the best human,
the game of chess is extremely popular and has had a huge surge particularly 
since 2020. Training of elite players is a full time job with many professionals
competing in well funded tournaments: on-line and over-the-board ones.

Chess presents an interesting ecosystem where open source and free access still has a 
strong hold that is almost unique. The relevant projects include a web site for playing 
and the best playing engine. In addition, and by and large, many games can be usually found 
easily, or be bought in large bundles relatively inexpensively. 
The Lichess website \cite{Lichess2026} is a free to play and open source site
with a code base available on \sw{github}. 
One of the best, if not the best engine for playing chess is \cite{Stockfish2026}
also developed as an open source project. The engine implements a lookahead of possible positions
using an estimated evaluation of possible future positions in a classic AI search algorithm.
Traditionally, engines will use metrics such as differential in material and positioning of pieces 
to numerically assign a preference towards one or the other player (black and white in 
the case of chess).

Around $2017$, neural networks made their presence felt in the world of chess 
when \sw{AlphaZero}(\cite{SilverD+2017}) by \sw{DeepMind} used a deep architecture with
reinforcement learning to beat the what at that time considered the best engines. Deep learning has since been 
central in chess playing engines. This was helped along with a popular 
book that analysed \sw{AlphaZero} games from a human perspective (\cite{SadlerMReganN2019}).
In addition, \sw{Stockfish} regained its dominance as a champion of engine tournaments 
by incorporating the output of neural networks for the evaluation of future possible positions.
Engines are considered to have a strength of over $3500$ (\sw{Stockfish} is estimated to be 
$3645$ in classical time control), whereas the best human has a current ELO (a numerical 
estimate of the strength of the player based on their wins and losses against others)
is $2840$ (\cite{Fide2026}). Despite this apparent dominance of engines, chess playing 
has become more popular. Contrary to what was expected, easy access to an excellent approximation of 
"objective truth" seems to spur players on, as well as increasing the general interest to the game.
The seamless transition between on-line and over the board playing 
along with the possibility of a variable time element are also likely contributing factors. 
Although the same game playing rules apply, games played at different time controls
provide very different experiences. The time range from games that last hours to bullet 
ones of duration of as little as $1$ minute with $1$ second increment.
Two factors that are often cited for the resurgent popularity over the past few years 
are the success of the The Queen's Gambit mini-series (\cite{FrankS2020}) and the 
lock-downs through the global pandemic of $COVID-19$.

For players that seek to improve their game, both at a professional and club-playing levels,
access to past games is a crucial part of training.
During general training the continuation of past games shows the most likely 
paths of games the player might have to face in future games along with the
likelihood of success for each move. When preparing for a tournament or specific 
opponent one would wish to be able to have access to the past games of the 
tournament participants or more often that of a particular opponent. 
Preparing against a specific opponent is common practice even at the lower 
amateur levels. At professional level a successful preparation is a crucial
step to success. Guessing correctly the path an opponent will follow means, 
that a player can navigate the early parts of the game with memorised good 
moves from past games and the output of their engine.

Given the popularity of on-line and over the board games, the number of 
available games to work with when preparing has exploded in number.
It is within living memory when players would only have access to past games through
published magazines and preparation in particularly opening via published books.
Currently, a huge resource for played games is \sw{Lichess} which provides all
the games played on the platform. Games can also be purchased in bulk from providers such as \sw{Chessbase}, \sw{chessgames.com}
and inexpensive sites such as the  \sw{Week in Chess} (\url{theweekinchess.com}).
We will use \pgn files from \elite \lichess database (\cite{elitelichess2022}).
The current availability of games make chess an interesting arena for trying data science approaches.

Despite the excellent positioning of open source chess communities, it is not immune 
to the global trend of commercialization. Open tools in high level programming 
environments have a critical role to play. Here we present a number of tools that are 
useful in manipulating chess boards and chess games in a logic programming environment.
Our suite of predicates forms an easy to install \swi (\cite{WielemakerJ+2012}) pack.
Packs are maintained by users, but are installable via a central manager and have a 
minimum of prerequisites they must abide to which makes them visible to the system
in a uniform manner.
The techniques and tools we present are not only pertinent to chess playing but promote
the role of logic programming in data science more generally.

\subsection{Paper Structure}
The remainder of the paper is structured as follows.
Section~\ref{sec:games} provides the building blocks for manipulating chess games within
\chessdb. Section~\ref{sec:store} discusses the storage of games to databases and 
especially considers experimental results of using a key-value database for storing 
chess position related information about past games.
Finally, we present a discussion and concluding remarks in Section~\ref{sec:concl}.


\section{Manipulating chess games}
\label{sec:games}

\Chessdb provides predicates for manipulating chess games in-memory and from/to file storage.
For in-memory manipulation we implement a chess board as a dictionary 
as implemented in \swi (\cite{WielemakerJ2026dicts}).
These are typical dictionary data structures where named arguments are stored holding a single piece of information each.
For reading games from files \chessdb implements a \sw{PGN} grammar that converts input
files in the portable game notation format to \prolog term structures.

\subsection{A board dictionary}
\label{sec:dict}

\begin{table}[t]
 \centering
 \caption{Starting position (LHS, Fig.~\ref{fig:queens}) dictionary.}\label{tbl:dict:start}


 \begin{verbatim}
 chess_dict_start_board(Board) :-
    Board = board{
            % cwk: castle white king-side; 1: available, 0: not-available
            cwk:1,cwq:1,cbk:1,cbq:1,
            eps:0,   % en passant square
            hmv:0,   % half moves clock since last take or pawn move
            fmv:0,   % full moves: played so far
            8:10,16:8,24:9,32:11,40:12,48:9,56:8,64:10,
            7:7 ,15:7,23:7,31:7 ,39:7 ,47:7,55:7,63:7 ,
            6:0 ,14:0,22:0,30:0 ,38:0 ,46:0,54:0,62:0 ,
            5:0 ,13:0,21:0,29:0 ,37:0 ,45:0,53:0,61:0 ,
            4:0 ,12:0,20:0,28:0 ,36:0 ,44:0,52:0,60:0 ,
            3:0 ,11:0,19:0,27:0 ,35:0 ,43:0,51:0,59:0 ,
            2:1 ,10:1,18:1,26:1 ,34:1 ,42:1,50:1,58:1 ,
            1:4,  9:2,17:3,25:5 ,33:6 ,41:3,49:2,57:4 ,
            0:0 % white move = 0; black = 1
    }.
 \end{verbatim}
\end{table}

Our dictionary stores the occupancy of each square in the numerical arguments $1-64$ travelling 
column-wise, $1$ being the $a1$ chess board square and $2$ being the $a2$ square. Each white piece 
is encoded in value $1-6$ and black pieces $7-12$. Further important information in the dictionary
are the castling rights for each rook (${cwk,cwq,cbk,cbq}$), where $cwk:1$ stands for it is still
possible for white to castle king side. That means that neither the white king nor the white rook 
on the king side have moved thus far in the game. $cwk:0$ would mean the white can no longer 
longer castle on the king side. 

\begin{figure}[t]


   \includegraphics[scale=0.9]{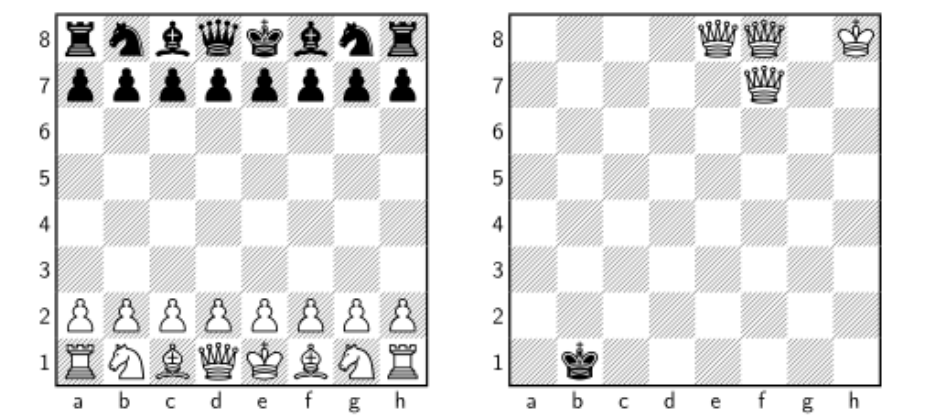}

    \caption{Left: Starting position. Right: An example of a notation where the departing square of a moving piece has
    to be fully qualified, as $3$ pieces pair-wise sharing column and row are able 
    to move to the landing square ($e7$).}
    \label{fig:queens}
\end{figure}

The dictionary also keeps a record of an en passant square if one is available on the current board.
This is via argument $eps$ which holds $0$ if such a square is not available, and $N$ ($1-64$) for the 
relevant square otherwise.
A maximum of a single en passant square is available at a board. That is the square behind a double square 
pawn move in the immediate previous move. So, if in the starting position (LHS,Figure~\ref{fig:queens} and Table~\ref{tbl:dict:start})
white plays $e4$, that is moves the pawn from $e2$ to $e4$ (a double square pawn move), at the next
move (black's turn to play), the square $e3$ is an en passant square. Signifying that were there
a black pawn at $d4$ or $f4$ then black would (in that move only) be allowed to take the $e4$ 
pawn by moving their pawn to $e3$. The white move $e4$ would change the value of $eps$ to $35$ 
which is the dictionary encoding for square $e3$.

Finally, the dictionary keeps information on the next to play (argument $0$), half moves clock ($hmv$) 
and full moves played so far. The half moves clock is a counter in (half moves, i.e. moves by each player
also referred to as ply in chess terminology) since the last capture or move of pawn. If $50$ full moves 
have elapsed since either of this has happened, either player can claim a draw.

The board for the starting position is shown in Figure~\ref{fig:queens} and the \prolog predicate defining the dictionary is shown 
in Table~\ref{tbl:dict:start}. Based on this representation \chessdb enacts moves in algebraic
form by changing values to the affected arguments. The algebraic notation is the standard notation
for recording chess games. Pawn moves simply record the square of the move if it is a forward move
(this uniquely identifies which pawn has moved) such as $e4$. When the pawn takes an opponent's 
piece or pawn, the column is given which again uniquely identifies the pawn, as any two pawn able to take 
material on the same opponent square will have to be in distinct columns (an example of the algebraic form 
of such a move is $exd5$).

Moves involving pieces are recorded with a starting capital letter denoting the piece involved
(kNight, Bishop, Rook, Queen and King). Typically, the landing square is given for the piece ($Nf3$).
If the move is a take a $x$ is inserted between the piece and the square ($Bxf3$).
When multiple pieces of the same type can land on the same square the column letter or row number 
is given after the piece letter ($Raxe4$ or $N3xe4$). 
In extremely rare cases and only when there are more than $2$ pieces of the same type on the board 
(only possible when pawns have been promoted), the exact square of the piece involved has to be given.
The example shown in Figure~\ref{fig:queens} is white's $95th$ move (\url{https://lichess.org/RBSi3ZQC#188})
with one of the three queens moving to square $e7$. In this instance the move was recorded as $Qf8e7$.

One of the trickiest parts of algebraic notation is when more than one pieces can occupy the same landing square, 
(typically that would $2$ pieces such as knights of rooks),
but one of them can not be moved due to a pin. A piece is pined when it is the only piece 
between its king and an attacking piece. In that case, the piece cannot be moved.
Our dictionary code implements full awareness of all legal moves and it
has been used to enact on the board moves from millions of games from PGNs containing legal notation (see Section~\ref{sec:experiment}).

\subsection{The PGN format}

\begin{table}[!ht]
 \centering
 \caption{Brief \pgn file example. Game $6$ from second match of world champion versus the Deep Blue Computer. New York 1997
 (\cite{KasparovGGreengardM2017})
 }
 \label{tbl:pgn}

\vspace{0.2cm}
\begin{verbatim}
[Event "IBM Man-Machine"]
[Site "New York, NY USA"]
[Date "1997.05.11"]
[EventDate "?"]
[Round "6"]
[Result "1-0"]
[White "Deep Blue (Computer)"]
[Black "Garry Kasparov"]
[ECO "B17"]
[WhiteElo "?"]
[BlackElo "?"]
[PlyCount "37"]

1.e4 c6 2.d4 d5 3.Nc3 dxe4 4.Nxe4 Nd7 5.Ng5 Ngf6 6.Bd3 e6 7.N1f3 h6 8.Nxe6
Qe7 9.O-O fxe6 10.Bg6+ Kd8 11.Bf4 b5 12.a4 Bb7 13.Re1 Nd5 14.Bg3 Kc8 15.axb5 cxb5
16.Qd3 Bc6 17.Bf5 exf5 18.Rxe7 Bxe7 19.c4 1-0
 \end{verbatim}
\end{table}

\pgn is a file format for storing chess games that is universally recognised by 
chess software. There are two main components in a game recorded in a \pgn file
as illustrated by the example in Table~\ref{tbl:pgn} (retrieved from \url{https://www.chessgames.com/perl/chessgame?gid=1070917} on 2026-02-03).
First, at the top part, there is information about the game, such as event, location
and date, along with names for the players and their strength ranking at the time
(Elo tags). Typically the result also appears here. The second part (separated by an empty line),
lists the moves in algebraic notation prepending with the number of the move.
The result might appear at the end of the moves, as in the example shown.

The notation allows for information relevant to specific moves to be added within curly brackets. For instance,
on-line platforms typically allow downloading of annotated games that, for instance, include extra information per move such as 
those produced by \lichess:
   \[
       \mbox{e4 \{ [\%eval 0.17] [\%clk 0:00:30] }\}
   \]
\noindent where \%eval is the engine evaluation and \%clk is the clock time before the move.


\noindent Consecutive games in a single file are separated by a single empty line.
\chessdb parses each game to a term of the form 
   \[
      pgn(Info, Moves, Res, Orig)
   \]

$Info$ is a pairs list containing all game-related information appearing on the first part of a game's \pgn representation,
and $Moves$ is a list of the algebraic form of single player moves.
$Res$ is the result of the game, and if the result appears both in the information section and at the end of the moves,
the code ensures there is agreement.
Finally $Orig$ is the verbatim original text appearing in the \pgn file.

\chessdb supports two methods for reading-in $pgn/4$ terms from files. First, all games are loaded in memory and the
user can manipulate them according to their task. Alternatively, games can be read one at the time, parsed, processed and
then forgotten. This mode of operation is useful when dealing with large files that would not fit in memory.
In this mode \chessdb can operate on extremely large data files- we don't expect there to be a limit
to the size as there is no cumulative demand on memory consumption.
When reading-in games incrementally, the user can either chose the default operation on each 
game, which is to store in a database (Section~\ref{sec:store}), or pass an arbitrary goal that processes one game at a time,
inclusive of an argument that passes possible accumulated information.
The pack program in \sw{pgn\_game\_lengths.pl} can be used to find the lengths of games \pgn files, using either 
in-memory loading, or by incremental processing (user needs to set the $incr(Incr)$ option accordingly).
Incremental processing is an important feature. In addition to other aspects, it can be used to break large files
before passing the partitions to popular graphical interface programs that cannot deal with large \pgn files.

%

\begin{figure}
    \includegraphics[width=1\textwidth]{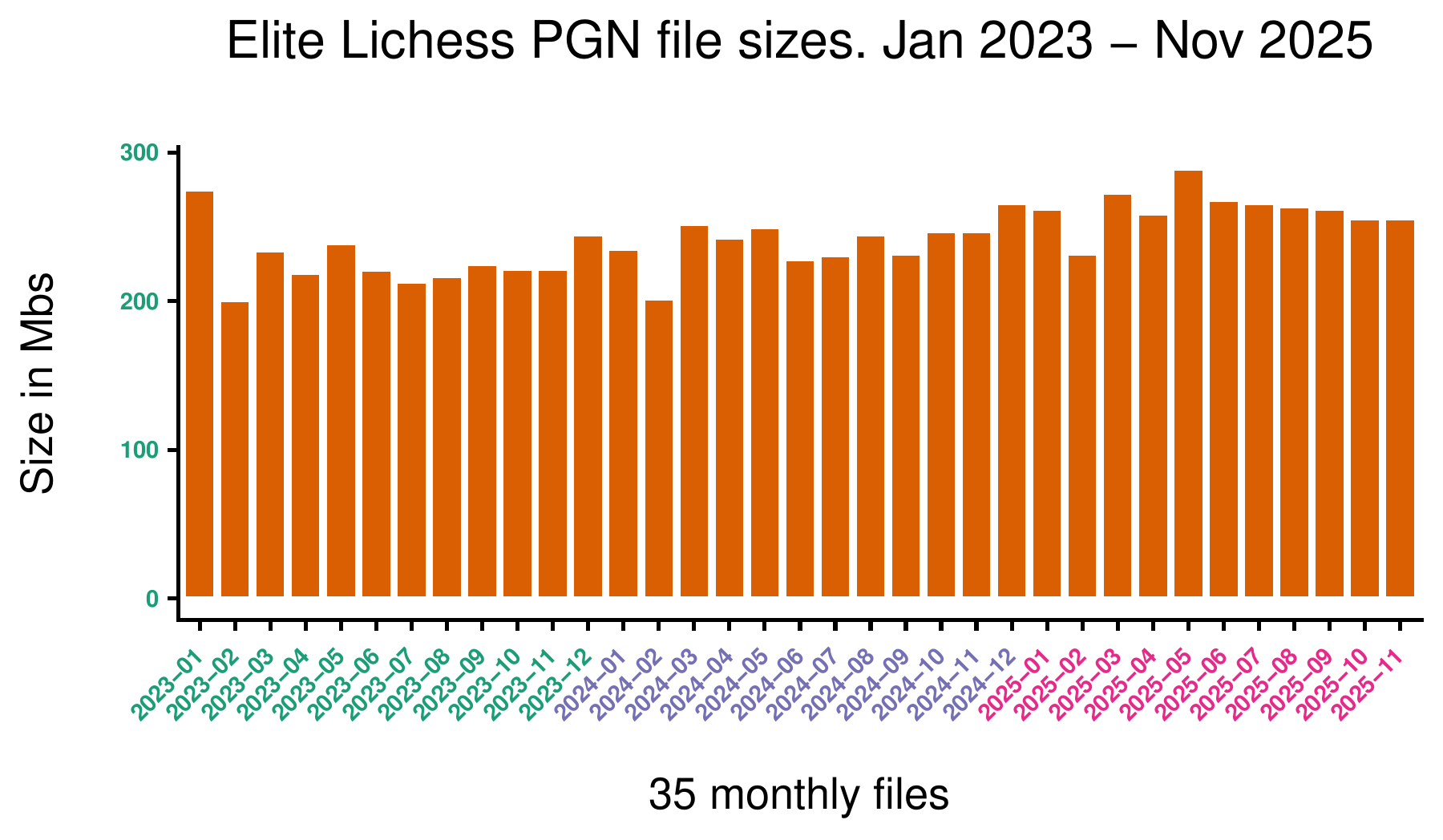}
    \caption{Size of files for the \elite \lichess \pgn files used in our experiments.}
    \label{fig:pgn:sizes}
\end{figure}

\begin{figure}
    \includegraphics[scale=0.4]{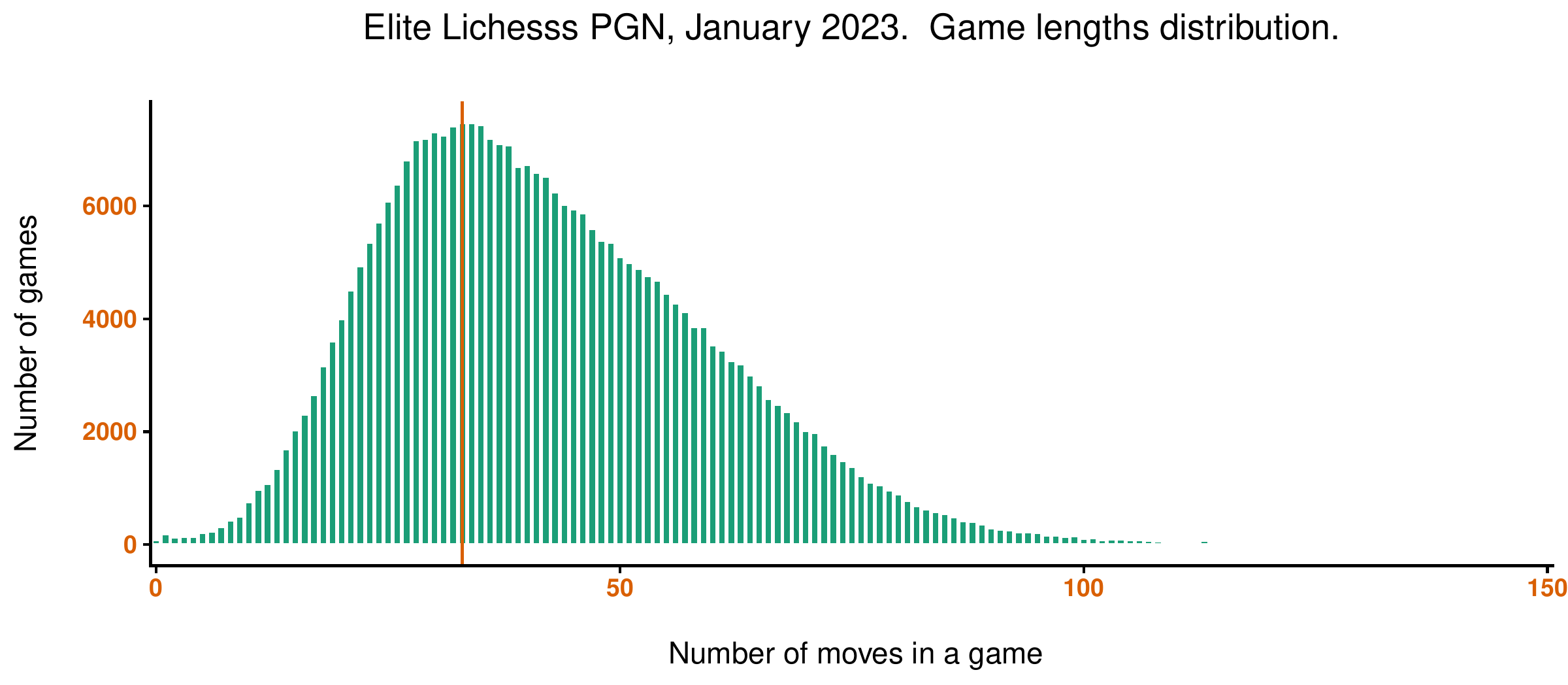}
    \caption{Game length distribution for the \elite \lichess \pgn file for Jan. 2023. The plot is curtailed at $150$ moves, removing $247$ games between $150$ and the maximum length of $301$. }
    \label{fig:glen:distro}
\end{figure}

\pgn files can be found from many sources, although it is unlikely a single source would fit any particular 
user's needs. A valuable resource of large number of games is \lichess which provide all games played in a single 
month on the platform, as a separate file (\url{https://database.lichess.org/}). Currently these 
are very large files. The compressed form (\sw{.zst}) has increased from $18MB$ in January $2013$ to a stable
size around $30Gb$ for the past $5$ years. It is worth noting that there was a marked increase in April $2020$. 
These files include all time controls, players of all abilities and all game variants. 
As we are only interested in the classical variant of chess and need a realistic, upper limit on the number of games,
we work with the \pgn files of the \elite \lichess database (\url{https://database.nikonoel.fr/}) which contains
monthly uploads since June $2020$.
We will experiment with the \elite files from $2023/01$ to $2025/11$. The size for each file
is between $200$ and $300$ Mbs (uncompressed) as shown in Figure~\ref{fig:pgn:sizes}.
The first file (January $2023$) contains $311,327$ games in total. The distribution of the length of the games in the same file
is shown in Figure~\ref{fig:glen:distro}. As it can be seen, although the most common length of game for that
month was $34$ (with $7470$ games, highlighted on Figure~\ref{fig:glen:distro} by the vertical dark orange line),
it is still the case that there a long tail in the distribution with a scattering of 
small number of games that last more than $100$ moves.
There are $1858$ games for which there are more than $100$ moves.
This is quite a large number, and upon inspection many of these games are on-line that played in non-competitive
spirit and include games that would never be included in a dataset for training players.
So, although \lichess is an excellent place for finding games, the quality even when only considering strong 
players, is not always guarantee, and it is likely serious players, will have collections of far smaller size, but
of much better quality. 
Here, we use the \elite set described above to create a test set of $10M$ games ($9,935,880$ to be precise).

In addition to free on-line platforms, game files from most tournaments can be downloaded although 
typically, and especially for long time formats, these will be a small number of games at any specific tournament.
Games can also be purchased in bulk from commercial providers such as \sw{Chessbase} and  \sw{chessgames.com}.
Finally, the website, \sw{Week in Chess} (\url{https://theweekinchess.com}) has been compiling tournament games since 
$1994$. The current collection contains $4$ million games and is available for a modest donation.


\section{Storing games to databases}
\label{sec:store}

While the \pgn format is able to record all game moves and associated information about details such as 
the location, date and player names, it is impractical for finding aggregate information about large number of games.
\chessdb provides code for storing games in databases which makes searching and working with games much 
faster. It supports two back-ends: \sw{proSQLite} and \sw{RocksDB}.

There is support for saving information in four tables, all implemented as independent databases:
\begin{description}
   \item{\textbf{info}} maps unique incremental game id to game information
   \item{\textbf{move}} maps game id to game moves
   \item{\textbf{orig}} maps game id to originial \pgn text
   \item{\textbf{posi}} position table where each position is mapped to possible continuation moves, along with
   winning frequencies for each move. Additionally store top $N$ games that have reached this position.
\end{description} 

The first three tables use a unique integer identifier to store the game related information (\sw{info})
the moves for a game (\sw{move}) and the original \pgn text verbatim (\sw{orig}).
The identifier is an internally database identifier that is increased for each new game.
The position table uses a unique encoding for each position to store aggregate information regarding this 
position from all games that seen this position. By keeping this information in this format, we
do not need to search for and visit all relevant games if we want access to this information.


\subsection{Unique position identifiers}

Converting a position to a unique identifier is used as the key value in mapping unique positions to
useful game continuation information.

Earlier versions of \chessdb, which were distributed but without being described in publication, 
included a naive implementation of representing positions as long integers.
These are supported in \sw{proSQLite} and since the databases we experimented 
with included only a few hundred games, efficiency was not an important issue. 
However, as multiple \pgn files with substantially more games in each file were used, performance
of both \sw{proSQLite} and the \prolog code became limiting issues.
The need for a far more economical representation that would also be supported by \rocksdb 
and re-writing of the code that generated, became a priority.

\begin{table}[ht!]
 \centering
 \caption{Starting position encoded as position table's key and the encoding of pieces and information about the game stage within piece nibbles.}\label{tbl:posi:start}

\begin{verbatim}

C3C3C3C3C3C3C3C3D17E21783179517B617C31792178D17E


chess_dec_hex(0, 0'0).    % en passant pawn
chess_dec_hex(1, 0'1).    % white pawn
chess_dec_hex(2, 0'2).    % white knight
chess_dec_hex(3, 0'3).    % white bishop
chess_dec_hex(4, 0'4).    % white rook
chess_dec_hex(5, 0'5).    % white queen
chess_dec_hex(6, 0'6).    % white king
chess_dec_hex(7, 0'7).    % black pawn
chess_dec_hex(8, 0'8).    % black knight 
chess_dec_hex(9, 0'9).    % black bishop
chess_dec_hex(10, 0'A).   % black rook
chess_dec_hex(11, 0'B).   % black queen
chess_dec_hex(12, 0'C).   % black king
chess_dec_hex(13, 0'D).   % white castling rook
chess_dec_hex(14, 0'E).   % black castling rook
chess_dec_hex(15, 0'F).   % black moving king 
   
 \end{verbatim}
\end{table}

Here, we follow a representation from an on-line article \cite{FiekasN2024}. According to the
source, a version of this was firstly introduced by the \cite{Stockfish2026} popular chess engine,
for their neural network training experiments.
In our scenario there is no need to store the half move clock and ply counters (1 byte each) or 
any variant data.
For our purposes a position is independent of how many moves it took to be reached (ply counter).
Strictly speaking, two identical positions that have different number of half move clock values are distinct
as they might affect the move decision. Half move clocks, enumerate the number of half moves (ply, each 
move contains a move for white and a move for black, that is $2$ plies) since the last take or last pawn move.
The importance of this is that when $50$ moves have elapsed the game is considered as drawn.
However, as we will argue later more extensively in Section~\ref{sec:experiment}, the positions table 
is only useful in earlier parts of the game, well before move $50$ or the rule of $50$ becomes an issue.

There are three important aspects we must preserve in addition to the positions and the pieces in each position.
First, that they have the same en passant square- if any, second  that they preserve
the same castling rights for black and white and third it must record the information about which side has the turn to move.
Otherwise, the positions are distinct, as there are different moves the players can chose from.
These items of information are encoded in the piece representation as there is space in terms of bits that
can be filled.
The representation contains two distinct components, the columns occupancy and the pieces occupying each 
marked square.
The occupancy contains a single bit of information ($0$ for a free square and $1$ for an occupied one) for
each square on the board. We group the bits in fours, thus needing $16$ groups for the whole board.
Following the occupancy, each present piece is encoded by half a byte as shown at Table~\ref{tbl:posi:start}.
We map each $4$-bit (nibble) of information to a hexadecimal digit to form a unique representation of 
positions to hexadecimal numbers.
The top of Table~\ref{tbl:posi:start} shows the representation of 
the starting position (one of the longest as all pieces are on the board).
The occupancy part of the starting position is given as:
\[
   C3C3C3C3C3C3C3C3
\]

The letter $C$ translates to binary $1100$ which shows the occupancy of the first $4$ rows of each of the columns
followed by its complement binary $0011$ which corresponds to hexadecimal $3$.
The occupancy is followed by the pieces occupying the signposted squares.
\[
   D17E21783179517B617C31792178D17E
\]

The term is almost symmetrical, with $D17E\mbox{ }2178\mbox{ }3179$ the $3$ symmetrical columns bookending 
the queen and king columns. The black rooks are mapped to $E$ as they are still available for 
castling. The black king is encoded by $C$ as it is white to move.
The same position where it is black's move, would not be identical and differ to the above in that 
the king will be encoded by $F$. Finally, as there is no pawn available to be taken by
en passant there is appears no $0$ in the pieces section. $0$ can be used here to encode this information, 
as unlikely the dictionary representation in Section~\ref{sec:dict} empty squares do not need to be
marked by a piece encoding. The occupancy bit matrix encodes the empty squares.


\subsection{proSQLite}

\chessdb supports \sw{{SQL}ite} via the \sw{proSQLite} pack (\cite{AngelopoulosNWielemakerJ2017}).
This interface library has been extensively tested and has a substantial number of downloads 
through the pack manager library, \sw{prolog\_pack}(\cite{Packs2026}).
It has been tested in a variety of contexts including in holding biological data (\cite{AngelopoulosNWielemakerJ2017}).
The \sw{SQLite} tables are as follows, where $+$ sign gives key fields and within the parentheses is the type.

\vspace{0.3cm}
   \begin{tabular}{rlcl}
   \hspace{0.3cm} & info(+Gid(int),+Key(atom),Value(atom))         & & orig(+Gid(int),Orig(atom)) \\ 
                  & move(+Gid(int),+Ply(int),Hmv(int),move(atom)) & & posi(+Pid(atom),Cont(atom))
   \end{tabular} 
\vspace{0.3cm}

The \sw{SQLite} is a very convenient back-end as it is zero configuration and the libraries are widely available
for all operating systems. It is particularly good for working with the info/3 table, allowing to search specific game
info quickly and conveniently. The library natively supports long integers. \sw{SQLite} is unsuitable for 
storing position tables from large datasets. We found that insert operations performance suffers after a few hundred thousand games.

\subsection{RocksDB}


RocksDB \cite{RocksDB2026} is an open source key-value high performance database management system.
The \swi pack \sw{rocksdb}(\cite{WielemakerJ2026rocksdb}) provides a convenient high level interface to the
\sw{RocksDB}, taking care of all the low level communication to the database. It has been available for a few years
and have been used in a number of projects, including the storage of \prolog clauses (\cite{WielemakerJ2022preds}).
The four tables are stored in four separate databases as follows:

\vspace{0.3cm}
   \begin{tabular}{rlcl}
      \hspace{1cm} & info(Gid(int64), KV(atom))  & \hspace{1cm} & orig(Gid(int64), Orig(atom)) \\
                   & move(Gid(int64), Mvs(atom)))&              & posi(Pid(atom), Cont(atom)) \\
   \end{tabular}
\vspace{0.3cm}

\rocksdb is less natural in capturing the game information as we collapse information tags and information values into a single atom.
However, it is particularly good in modeling the position table, with the position hexadecimal as the atomic key of the database and 
the continuation atom as the value of the database. The database extends the range of number of games that can be stored to millions 
in comparison to the \sw{SQLite} interface.
We detail below some experimental results with position \rocksdb databases created with \chessdb.

\subsection{Experimental results for position table in a key-value database}
\label{sec:experiment}

To test various performance aspects of the positions \rocksdb database we downloaded the monthly \elite \lichess database uploads from
\url{https://database.nikonoel.fr/} ranging from January $2023$ to November $2025$. The games are from \lichess, trimmed for standard 
chess games played by members with Elo of at least $2500$ versus $2300$ which was then updated from 
These contain all time controls apart from bullet games ($2$ minutes or less).
The size of files is stable around $65Mb$ containing similar number of games ($311,327$ in the case of Jan 2023) and lines ($8,362,602$ in Jan 2023 set).
The distribution of games for the $35$ input files is shown in Figure~\ref{fig:pgn:sizes} while Figure~\ref{fig:glen:distro} 
plots the distributions of length of games within one of the \pgn files.

Position tables are useful when training or analysing the opening part of the games, when there is a chance that a number of games have 
been played from that specific position. Training players can at each position check the played continuations along with
the results they led to and links to $N$ games that have seen the position. The chances of a meaningful number of games to have been seen at a position 
diminishes rapidly as the game progresses. It is often the case that databases use a hard limit ($M$) of move depths to limit the data stored.
\lichess appears\footnote{as far as we can see, these are undocumented, but we postulate from observing the web-interface} 
to be using $M=25$ and $N=8$ split across $4$ most recent games and $4$ games played by the best players.

We performed experiments on a single month game set, to see the effect of the depth limit of $25$. 
For this we used the January $2023$ \elite dataset, this as well as the other sets include
a number of atypical games, such some that last for $300$ moves.
Creating positions for unrestricted move depth created on average $68$ positions per game.
There was no significant difference in performance as the increase was not significant with respect to the database capabilities.
For the remaining of the experiments we used a maximum depth of $M=25$. 
We will also use $N=10$ to store the game ids to the $10$ games played by the strongest players.
We create a single database of $10M$ games from the $35$ files as described above. 
We ran two experiments. In the first we build the database in one go, while in the second we restart inserting 
to the databases at three points: $7$,$13$ and $25$.

\begin{figure}
    \centerline{\includegraphics[scale=.42]{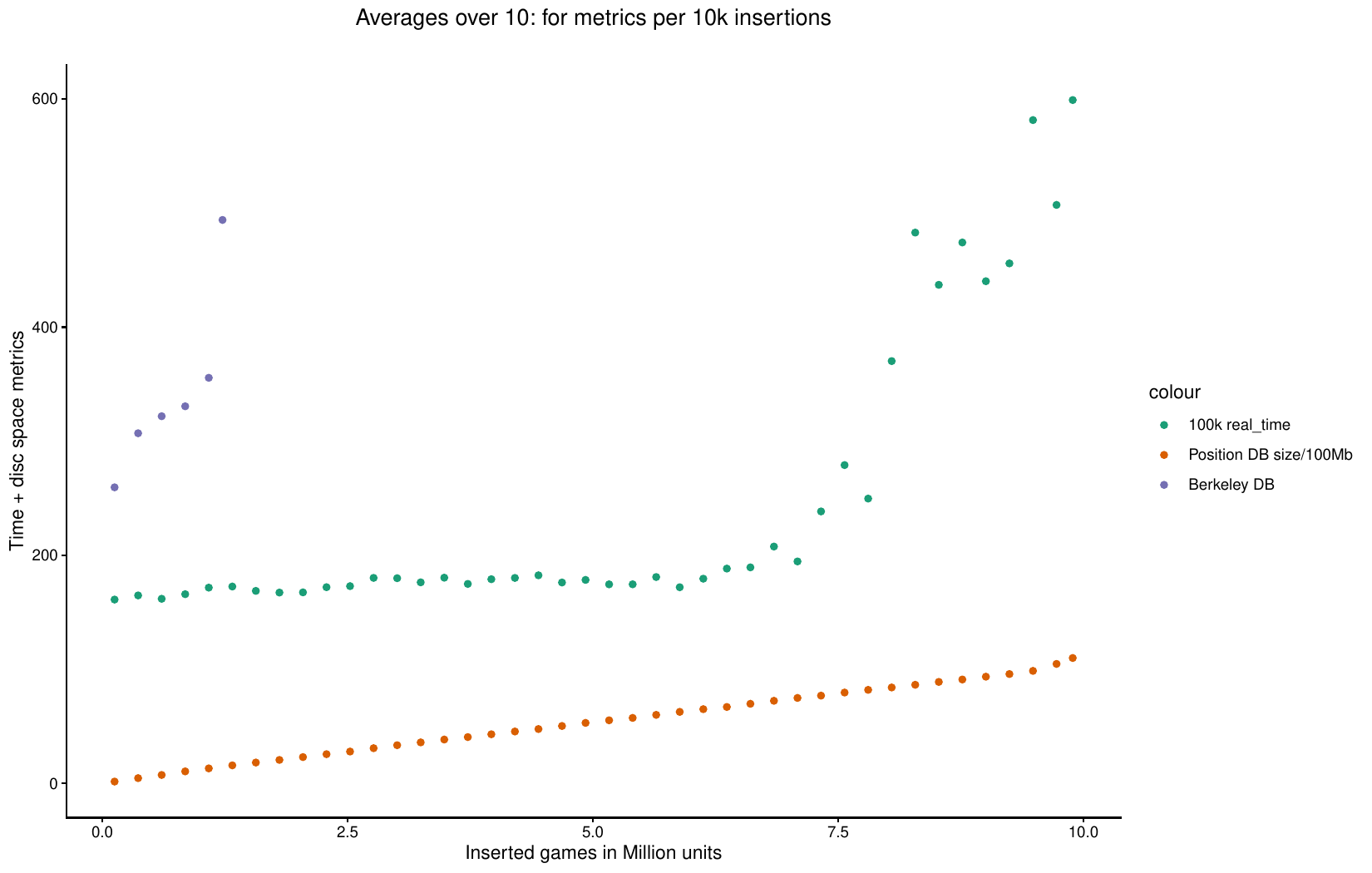}}
    \caption{Statistics from single pass experiment.}
    \label{fig:stats}
\end{figure}

We identified $4$ stages of performance. The database performs extremely well up to $3M$ games with below $3$ minutes with some margin per $10,000$
games. In the broad range to $6M$ games, the performance is just over $3$ minutes. From $6M$ to $9M$ the performance progresses steadily to $5$ minutes
and for the last million it increases further to $8-10$ minutes. These timings are in actual time passed.
Using the \sw{real\_time} key in predicate \pred{statistics}{2}, we plot the insertion times for the single pass experiment at Figure~\ref{fig:stats}.
Each dot is the average of $10$ segments each inserting $10K$ games.

It is worth noting that restarting insert operations to a large database seem to incur an initial cost. 
In our experiments the $second$ and $third$ restart showed a temporary slow down of $3$ hours in the first case,
until the performance of insertions stabilised, and of $5$ hours in the third restart.
Storage usage grows linearly to the games store with $10M$ games taking $11.2Gb$ (bottom line in Figure~\ref{fig:stats}).

We also ran the same experiment on \sw{Berkeley DB} which is also a key-value database. \swi provides an interface to the database which ships
with the main system. The performance of this database system starts at marginally worst than \sw{RocksDB}. However, it deteriorates
within the $1$ million insertions and becomes unusable by $1.3$ million iterations (reaching $1$ hour for some $10k$ insertions, and consistently requiring more then
$30$ minutes after this point).

Experiments were performed on a dated Linux desktop with $16Gb$ of memory, with an \sw{AMD A8-6600K} processor
(4 Cores, 4 Threads, 1 CPU, released Jun 1st, 2013).
We used default \sw{RocksDB}, pack(\rocksdb) and \sw{SQLite} settings.
Our experiments were about finding the limits of the approach rather than recreating a realistic environment.
It is unlikely that in a envisaged context such as individual player, coach or team using the software 
they will be $10M$ games that can be meaningfully bunched together, particularly in the context of position tables. 
It is far more likely that the total number of games trusted will be in total of much smaller size, while they
will also be bands of separation in terms of time controls and player expertise. For instance, \lichess  
gives an option between a Masters curated (unpublished) database of $1.3M$ games or combinations of their on-line
games split in $6$ time controls and $9$ average player ratings.

\section{Discussion and Conclusions}
\label{sec:concl}


\Chessdb is easily installable via the \swi (\cite{WielemakerJ+2012}) pack manager, or directly from source and is available on github\footnote{
\url{https://github.com/nicos-angelopoulos/chess_db}}.
The \sw{SQLite} database back-end requires \sw{SQLite} and packs \sw{proSQLite} and \sw{db\_facts} which are easily installable in all operating systems.
The \rocksdb engine requires \sw{RocksDB} and pack \sw{rocksdb}. The latter will in \sw{Linux} install the \sw{RocksDB} from sources in a 
lengthier and more technical task taking over a gigabyte of space most of which can be reclaimed.
At load time \chessdb decides according to user instructions, to load one of two supported back-ends or a hybrid one, which supports both at
a small cost of performance.  The features detailed here refer to version \chessdb $v1.2$ and later.

The library provides predicates for (a) parsing \pgn files (\pred{pgn}{2}), (b) processing \pgn files, especially for adding games to databases
(\pred{chess\_db}{2}), (c) working with chess databases (\pred{chess\_db\_connect}{1}, \pred{chess\_db\_game\_info}{3}) and (d) working with
chess dictionaries (\pred{chess\_dict\_start\_board}{1}, \pred{chess\_dict\_move}{5}).
\Chessdb is an open source approach to declarative data science that fits well with the, still strong, persisting ethos of openness 
and free access to the game and information about the game in the community.
It can be used by serious or professional players and coaches of individuals and teams.
Chess is one of the few globally successful arenas where the paradigm of open source still holds strong.

Unlike \lichess which uses \sw{MongoDB} and targets multi-player web interfaces here we explored tools 
that are more appropriate for personal and team usage. The software would also be useful to data scientists
working with chess data.  The use of database technologies that require less technical knowledge is also
an important feature. The declarative nature of \chessdb allows for high level programming analytics.
In addition, it is complementary to other \swi tools such as graph plotting via \sw{Real} (\cite{AngelopoulosN+2013,AngelopoulosN+2016a}).
Among other scenarios, \chessdb can also be used to narrow games and create smaller \pgn files from large collection that can 
be viewed with chess visualisation programs such as \sw{en-croissant}, \sw{Scid}, \sw{Chessbase} and \sw{ChessX},
which are unable to digest large files.

Here we also formalised things that are often embedded in software implementations and we provide a solid basis for future work
with chess games within Logic Programming environments.
One of the priorities for future work includes connecting \chessdb to graphical interfaces.
While in term of efficiency improvements, there are two likely avenues in move disambiguation and creating the \prolog terms 
for each game read from a \pgn. Currently, we use a two sweep approach, in order to keep the full
original text as well as a parsed term representation. Collapsing this to a single step process
is likely to provide some marginal gains.

\section{Acknowledgments}
We are indebted to the referees for the incisive comments. In addition to referee $1$ for their meticulous reading and associated corrections and to
referee $2$ for the chess related comments.

\bibliographystyle{eptcs}
\bibliography{final}

\end{document}